\documentclass[showpacs,nofootinbib,twocolumn,secnumarabic,amssymb, nobibnotes, aps, prc, nolongbibliography]{revtex4-2}
\usepackage{xcolor}

\usepackage{graphicx}  % needed for figures
\usepackage{dcolumn}   % needed for some tables
\usepackage{bm}        % for math
\usepackage{amssymb}   % for math
\usepackage{amssymb,amsmath,amsfonts} % for math
\usepackage[breaklinks=true,debug=true]{hyperref}
\usepackage{braket}    % for QM notation
\usepackage{multirow}
\usepackage{adjustbox}
\usepackage{gnuplottex}
\usepackage[T1]{fontenc}
\usepackage[utf8]{inputenc}
\usepackage{dsfont}
\usepackage{tikz}
\usepackage{filecontents}
\usepackage{pgfplots}
\usepgfplotslibrary{groupplots}
\pgfplotsset{compat=newest}
\usepgfplotslibrary{units}

\newcolumntype{M}[1]{>{\centering\arraybackslash}m{#1}}

\begin{document}

%\widetext
%\leftline{Version 01 as of \today}
%\leftline{To be submitted to PRC}

\title{Realistic evaluation of Coulomb potential in spherical nuclei and test of the traditional approach}

\author{L.~Xayavong}
\email{xayavong.latsamy@yonsei.ac.kr}
\author{Y.~Lim}
\email{ylim@yonsei.ac.kr}
\affiliation{Department of Physics, Yonsei University, Seoul 03722, South Korea}

\vskip 0.25cm  
\date{\today}

\begin{abstract} 

A realistic evaluation of Coulomb potential has been made for some selected nuclei using the available
model-independent data for the charge density and the recent development of Coulomb energy-density functional. 
Within the Woods-Saxon potential as a nuclear component, we are able to quantify the differences in proton single-particle energies due to the differences from the model-independent data of the uniform distribution, the two-parameter Fermi function, as well as the charge density obtained from a microscopic Hartree-Fock calculation using the effective Skyrme interaction.  
The obtained energy differences are generally small in magnitude, namely about 100~keV or less if the parameters of the charge density models are appropriately determined. 
Considerable larger differences appear when the last occupied state is highly filled and, at the same time, has a small orbital angular momentum. 
Sulfur isotopes ($Z=16$) are a perfect example of these nuclei. 
Unfortunately, despite its simplicity, the uniform distribution cannot be used for evaluating the Coulomb exchange term within a well-established method because it is not differentiable at the surface of a nucleus. 
Traditionally, the missing of exchange term is corrected for by simply excluding the last-proton contribution to the direct term. 
We also investigate this approach and find that its effect is simply an introduction of the factor $(Z-1)/Z$ into the Coulomb direct term. 
From medium to heavy nuclei (typically beyond the $sd$ shell) 
%%
%%
%%%\tcb{Q: don't we need the principal quantum number for $sd$ shell, such as $1sd$ or $2sd$. I don't know the convention for this so I'm wondering the prospective readers understand this $sd$ naturally}, 
%%
%%\tcg{Normally we use `$sd$ shell', but some other shell-model groups prefer `$1s0d$ shell' (they count the radial quantum number from 0). Both conventions are acceptable. } YH;-> Thanks greaT!
%%
the resulting proton levels are 300-800~keV higher than those obtained with the exact Fock term. 
The result for lighter nuclei tends to be opposite because the factor $(Z-1)/Z$ decreases rapidly towards the limit of $Z\to 1$. 
Therefore, this traditional approach should be avoided for a precision nuclear structure calculation. 

\end{abstract}

%\pacs{21.60.Cs, 23.40.Bw, 23.40Hc, 27.30.+t}
\maketitle

\section{Introduction}

The Coulomb repulsion between protons is the best known part of nuclear Hamiltonian. Its action results in a large class of nuclear properties, including the departure from the $N=Z$ line of the stability valley or the fission phenomenon which yields a clean-cut limitation to the nuclear size, and some specific properties of the $\beta$, proton and $\alpha$ radioactivities. As the main actor in the violation of the isospin symmetry, the Coulomb interaction is also responsible for various nuclear structure phenomena, in particular it induces isospin mixing in nuclear states, displacement energy between members of isobaric multiplet and isospin forbidden transitions~\cite{OrBr1985,AUERBACH1983273,LAM2013680,PhysRevC.103.024316}. 
In addition, the correction due to the isospin-symmetry breaking has became a limiting factor in the low-energy tests of the electroweak sector of the Standard Model via nuclear $\beta$ decays~\cite{HaTo2020}. 
Therefore, in a microscopic model of the nuclear structure, it is of paramount importance to properly take the Coulomb interaction into account. 

Within the independent-particle framework, the $A$-nucleon problem is simplified to $A$ single-nucleon problems starting either from an effective nucleon-nucleon interaction~\cite{PhysRevC.5.626,BENNACEUR200596} or an effective potential~\cite{WS1,SWV}. 
Therefore, the resulting nuclear wave function is simply an antisymmetric product of individual wave functions or the so-called Slater determinant. 
As is well known, this simplified model itself has a limited application. In principle, the self-consistent spherical Hartree-Fock (HF) mean field is only appropriate for a closed-shell system, 
whereas the phenomenological one is usually optimized for single-particle or single-hole states such as low-lying states of nuclei in the vicinity of a closed-shell core. Nevertheless, the independent-particle potential serves as the basic part of the full nuclear Hamiltonian and therefore is the key for success for all nuclear many-body approaches. 
In some specific applications, it is mandatory to include the Coulomb contribution within the one-body part, especially in the shell model for which the valence spaces are too small to produce all significant configurations induced by the Coulomb interaction. 
As an example, the shell-model description of the superallowed $0^+\to 0^+$ Fermi $\beta$ decay would not agree with the Standard-Model predictions, unless the harmonic oscillator basis is replaced with realistic radial wave functions~\cite{ToHa2008,XaNa2018,XaNa2022}. In particular, the radial mismatch between the initial and final states is greatly enhanced when the mother nucleus is weakly bound. 

In nearly all calculations within the phenomenological Woods-Saxon (WS) potential~\cite{WS1,WS2,WS3,SWV,doi:10.1142/3530}, the Coulomb repulsion is accounted for using the approximation of a uniformly charge distributed sphere of radius $R_C$: 
\begin{equation}\label{eq1}
V_C(r) = (Z-1)e^2 \left\{
\begin{array}{ll}
\displaystyle \frac{1}{r}, & r>R_C \\[0.1in]
\displaystyle \frac{1}{R_C}\left( \frac{3}{2} - \frac{r^2}{2R_C^2} \right), & \text{otherwise,}
\end{array}
\right.
\end{equation}
where $R_C$ is usually taken as $R_C=r_0(A-1)^\frac{1}{3}$ with $r_0\approx 1.26$~fm~\cite{SWV}. 
Alternatively $R_C$ can be extracted from the charge radius $R_{ch}$ via~\cite{Elton}
\begin{equation}\label{eq2}
\displaystyle R_C^2 = \frac{5}{3} R_{ch}^2 - \frac{5}{2} \sum_{i=1}^3 \theta_i r_i^2 - \frac{5}{4}\left( \frac{\hbar}{mc} \right)^2 + \frac{5}{2} \frac{b^2}{A} 
\end{equation}
where the nuclear oscillator length parameter is given by $b^2\approx A^\frac{1}{3}$~fm$^2$. An improved parameterization of $b^2 $ can be found in Ref.~\cite{Kir2006}. 
The last three terms on the right-hand-side of Eq.~\eqref{eq2} account for the internal structure of proton where $\sum_i \theta_i r_i^2=0.518$~fm$^2$~\cite{BABrown_1979}, Darwin-Foldy term ($\hbar/mc=0.21$~fm) and center-of-mass (COM) motion, respectively. 
It should be noted that a further modification was made in the construction of Eq.~\eqref{eq1}, namely the contribution of the last proton was excluded so that the potential in Eq.~\eqref{eq1} is proportional to $Z-1$ instead of $Z$ as seen in classical electromagnetism. This exclusion of the last proton can also be considered as the correction of missing of the Coulomb exchange potential in the uniform charge approximation.
However, this approach for the self-interaction correction has not been rigorously checked for a microscopic nuclear structure calculation. 
%%
%%\tcg{A considerable modification was made to this paragraph, could you please read it again ?}

In general, the Coulomb potential can be derived from the two-body Coulomb interaction using the variational principle. 
Its direct part is basically a known functional of charge density, while its exchange counterpart can be treated with a great precision using a local density approximation~\cite{PhysRevC.99.024309}. 
Therefore, the Coulomb potential can be determined in a self-consistent manner by minimizing the total energy as in the HF theory~\cite{PhysRevC.5.626}, 
or evaluated independently from nuclear component using charge density data from external sources. 

In the present work, Eq.~\eqref{eq1} is investigated in various aspects. 
The validity of the uniform charge distribution which is the central pillar of Eq.~\eqref{eq1} is checked within the framework of the phenomenological WS mean field and the model-independent data deduced from electron scattering experiments~\cite{DEVRIES1987495}. 
The errors due to the above-mentioned 
approach for the self-interaction correction
are separately quantified 
through the comparison to the results obtained with the exact Coulomb exchange functional. 
%%
%\tcg{Here I replaced `double-counting correction' with `self-interaction correction' and also made an extension to the previous sentence. Could you please check it ?} 
%%
Moreover, we discuss the difficulty in evaluating the Coulomb exchange term when a uniform distribution is assumed or when the model-independent data are used instead. 
As a sensitivity study, we also consider the two-parameter Fermi (2pF) function and the microscopic Skyrme HF calculation as an alternative model for the charge-density distribution. 
Our calculations are performed for a wide variety of nuclei and covers a mass range between $A=16$ and $209$. 
There includes two closed-shell $N=Z$ nuclei ($^{16}$O and $^{40}$Ca), two closed-shell $N\ne Z$ nuclei ($^{48}$Ca and $^{208}$Pb), 
two closed-subshell nuclei ($^{28}$Si and $^{32}$S) and four opened-shell nuclei ($^{58}$Ni, $^{205}$Tl, $^{206}$Pb and $^{209}$Bi). 

%This paper is divided into six sections. 
The paper is organized as follows. 
The standard parameterization structure of the WS potential is described in the section \ref{sec:nucpot}. 
Section \ref{func} reviews the density functional forms of the Coulomb direct and exchange terms as employed in the self-consistent mean-field theory. 
The selection of input charge densities and discussion of their properties are given in the 
section \ref{input}
In the section\,\ref{sec:coulombpot}, a comparative test of Coulomb potential for different charge density models and different functional forms is carried out. 
The summary and conclusion are given in the last section\,\ref{sec:con}. 

\section{Nuclear potential}\label{sec:nucpot}

In this work, the phenomenological WS potential is selected for the nuclear component of our independent-particle Hamiltonian. As a standard parameterization structure, this potential consists of a central, spin-orbit, isospin-dependent and Coulomb term, namely  
%%
%%\tcg{Here I added `term'}
%%
\begin{equation}\label{ws}
\begin{array}{ll}
V(r) =& \displaystyle V_0f_0(r) - V_s \left(\frac{r_s}{\hbar}\right)^2\frac{1}{r}\frac{d}{dr}f_s(r)\braket{\vec{l}\cdot\vec{\sigma}} \\[0.1in]
     &+ V_{sym}(r) + (\frac{1}{2}-t_z)V_C(r), 
\end{array}
\end{equation}
where $t_z$ is the isospin projection of the nucleon with the convention of $t_z = \frac{1}{2}$ for neutron and $-\frac{1}{2}$ for proton. 
The functions $f_i(r)$ are defined as 
%%
%%\tcg{Here I avoided the term `form factor'}
%%
\begin{equation}\label{fi}
f_i(r) = \frac{1}{1+\exp\left( \frac{r-R_i}{a_i} \right)}, 
\end{equation}
with $i=0$ or $s$ denoting either the central or spin-orbit terms. The restriction of $a_0=a_s$ is usually adopted because of lack of experimental constraints. On the contrary to the assumption $a_0=a_s$, 
a smaller spin-orbit radius $(R_s< R_0)$ was suggested because the two-body spin-orbit interaction has a shorter range~\cite{doi:10.1142/3530}. 
For example, $R_s/R_0=0.921$ was obtained for the Seminole parameterization given in Ref.~\cite{SWV}. 

The expectation value $\braket{\vec{l}\cdot\vec{\sigma}}$ appearing in the spin-orbit term can be written as
\begin{equation}
\braket{\vec{l}\cdot\vec{\sigma}} = \left\{
\begin{array}{lll}
l & \text{if} & j=l+\frac{1}{2} \\[0.1in]
-(l+1) & \text{if} & j=l-\frac{1}{2}. 
\end{array}
\right.
\end{equation}

In order to preserve the fundamental symmetries, the phenomenological effective potential such as WS is normally treated as a nuclear mean field created by the core of $(A-1)$ nucleons. The exclusion of last nucleon contribution also serves as a self-interaction correction due to the missing of exchange terms as discussed for the Coulomb potential in the introduction. 
%%
%\tcg{In the present context, I think the term `self-interaction correction' is more precise than `double-counting correction'} -> Ok, great!
%%
For these reasons, the WS radii are usually parameterized as a function of $(A-1)$ instead of $A$ namely $R_i=r_i(A-1)^\frac{1}{3}$. Furthermore, if we neglect for instance the internal structure of the core of $(A-1)$ nucleons, 
the nucleus can be regarded as a system of two point-like particles. 
Within this simplified picture, the COM correction for the WS Hamiltonian can be easily implemented by replacing the nucleon mass $m$ in the kinetic energy term of the radial Schr\"{o}dinger's equation with the reduced mass $\mu$ defined below,  
\begin{equation}
\mu = m\frac{(A-1)}{A}.
\end{equation}
This COM correction is fully validated at large separation where the core structure contribution is negligible. A further discussion of the COM corrections is given in Refs.~\cite{Xthesis,XaNa2022}. 

Most nuclei have a different number of protons and neutrons. 
Apart from the Coulomb repulsion, different number of neutrons and protons in nuclei causes an extra shift between neutron and proton potential depth. 
As an experimental evidence, nuclei with $N=Z$ tend to have a greatest binding energy comparing with the other possible configurations of protons and neutrons. 
In practice, this effect is commonly accounted for by adding the symmetry term expressed below, 
\begin{equation}
V_{sym}(r) = 2t_z V_1 \frac{(N-Z)}{A} f_0(r). 
\end{equation}
%%
%%
%% No indent..
In this work, we neglect the symmetry term contribution to the spin-orbit coupling~\cite{XU2013247} for simplicity. 
Some further additional terms or slightly different parameterization structure can also be found in literature. For example, a more fundamental form of the isospin-dependent term was proposed by Lane in Ref.~\cite{PhysRevLett.8.171}. Study of such variations of the nuclear component is outside the scope of this work. 

The Coulomb potential $V_C(r)$ can be evaluated within various different methods, it will be separately described in the following sections. 

Basically, the WS potential cannot be used for the total binding energy since it is not based upon a
specific effective two-body interaction. Normally the WS parameters ($V_0$, $V_s$, $r_0$, $r_s$, $a_0$, $a_s$, $V_1$, $R_C$) are chosen for a best fit of nuclear single-particle energies and nuclear charge radii. 
The set of WS parameter values named BM$_m$ as listed in Table I of Ref.~\cite{XaNa2018} will be used in the present study. 

\section{Coulomb potential as a charge density functional}\label{func}

Before discussing the details of our study, it is instructive to describe in this section the basic formulas and some general properties of the Coulomb potential. 
We recall that, according to the self-consistent HF theory~\cite{PhysRevC.5.626}, the Coulomb contribution to the mean field consists of a direct and an exchange term
which can be symbolically written as
\begin{equation}
V_C(r) = V_{dir}(r) + V_{exc}(r). 
\end{equation}
Note that the spherical symmetry is assumed throughout this paper. 
The Coulomb direct term for a spherically symmetric nucleus, after integrating out the angular variables, is reduced to
\begin{equation}\label{dir}
V_{dir}(r) = 4\pi e^2 \left[ \frac{1}{r}\int_0^r x^2 \rho_{ch}(x)\, dx + \int_r^\infty x\rho_{ch}(x) \,dx \right]. 
\end{equation}
One can observe here that if the charge density $\rho_{ch}(r)$ is constant inside the radius $R_C$ and vanishes elsewhere, the expression~\eqref{dir} will return the potential in Eq.~\eqref{eq1} except that it would be proportional to $Z$ instead of $Z-1$ for the reason discussed in the introduction. More details on the uniform charge distribution are given in subsection~\ref{ax}.

Since the Coulomb force has an infinite range, the Coulomb exchange term is nonlocal in coordinate space and thus much harder to calculate 
especially within the self-consistent mean-field framework.
%%
%%\tcg{Here I added a sentence. Could you pleae check it ? ok}
%%
To avoid this inconvenience, a local density approximation is usually employed. The popular one is that invented by Slater~\cite{PhysRev.81.385} with which the Coulomb exchange term is given by a function of the charge density 
\begin{equation}\label{Slater}
V_{exc}^{Sla}(r) = -e^2\left[ \frac{3}{\pi}\rho_{ch}(r) \right]^\frac{1}{3}.
\end{equation} 
The quality of this approximation was checked against the exact calculation~\cite{PhysRevC.63.024312} for a number of spherical nuclei from $^{16}$O to $^{310}_{126}$Ubh. The proton energy levels found in Ref.~\cite{PhysRevC.63.024312} were underbound by 100 to 550~keV for occupied states and overbound by 100 to 200~keV for unoccupied states, compared with those obtained with the exact Fock term. %\tcb{ Question : underbound and overbound compared with the one without the Coulomb exchange?} %\tcg{no, compared with the levels obtained with the exact Fock term. To be clearer I added this %sentence into the above paragraph. Could you please take a look ?}

It was recently demonstrated that the Coulomb energy density functional built by using the generalized gradient approximation (GGA)~\cite{PhysRevC.99.024309} produces almost the same accuracy for the total energy as the exact treatment of the Fock term while the numerical price is still the same as that of the Slater approximation. The Coulomb exchange term derived from the GGA depends not only on the charge density but its gradient with respect to radial distance, namely 
\begin{widetext}
\begin{equation}\label{GGA}
V_{exc}^{GGA}(r) = V_{exc}^{Sla}(r) \left\{ F(s) - \left[ s + \frac{3}{4k_Fr} \right]F'(s) + \left[ s^2 - \frac{3\rho''_{ch}(r)}{8\rho_{ch}(r)k_F^2} \right] F''(s) \right\}.
\end{equation}
\end{widetext}
where $\rho''_{ch}(r)$ denotes the second derivative of $\rho_{ch}(r)$ with respect to $r$, whereas $F'(s)$ and $F''(s)$ denote the first and second derivatives of $F(s)$ with respect to $s$, respectively.
Note that $F(s)$ is the enhancement factor introduced for the Coulomb exchange potential in GGA. 
Following Perdew-Burke-Ernzerhof~\cite{PhysRevLett.77.3865}, $F(s)$ is parameterized as 
\begin{equation}
\displaystyle F(s) = 1 + \kappa - \frac{\kappa}{1+\mu s^2/\kappa}, 
\end{equation}
where the two parameters $\kappa$ and $\mu$ have recently been revised for nuclear physics applications by Naito {\it et al.}~\cite{PhysRevC.99.024309} and the best-fit values were found to be 0.804 and 0.274, respectively. The function $s$ denotes the dimensionless density gradient 
\begin{equation}
s = \frac{|\boldsymbol{\nabla} \rho_{ch}(r)|}{2k_F\rho_{ch}(r)}, 
\end{equation}
where $|\boldsymbol{\nabla} \rho_{ch}(r)|$ is the norm of $\boldsymbol{\nabla} \rho_{ch}(r)$ and the Fermi momentum is defined as $k_F=[3\pi^2\rho_{ch}(r)]^\frac{1}{3}$. We remark here that if $\rho_{ch}(r)$ is a slowly varying function of $r$, its gradient will be vanished ($s=0$) and thus $F(0)=1$, then the Coulomb exchange term in Eq.~\eqref{GGA} will be reduced to the Slater approximation~\eqref{Slater}. Considering the general behavior of charge density in a nucleus, 
the Coulomb exchange potential from the GGA is mostly affected in the nuclear surface region where the charge density gradient is peaked.

The contribution of higher order electromagnetic effects such as the vacuum polarization and the Coulomb spin-orbit can also be included. However, both of them were found to be completely negligible~\cite{XaNa2022}. 

\section{Input charge densities}\label{input}

According to the formalism reviewed in the previous section, the charge density is regarded as the fundamental ingredient with which the Coulomb potential is determined. 
In the following list, we briefly describe the conventional methods for deducing charge density from electron scattering or the so-called model-independent analyses. 
The data obtained from this source is used as the reference for our comparative study.   
The frequently-used hypothetical and phenomenological models as well as the microscopic self-consistent mean-field calculation of the charge density are also discussed. 
The validity of these theoretical models as the input of the Coulomb potential is investigated in the next section.

\subsection{Model-independent analyses}

The charge form factors for some stable nuclei were accurately measured by electron scattering. 
These data are usually analyzed within two different model-independent approaches to extract numerical values for the charge density as a function of $r$ namely the Fourier-Bessel\,(FB)\,\cite{DREHER1974219} and the sum-of-Gaussians\,(SOG)\,\cite{SICK1974509} expansion. 
In the former approach, the charge density is expanded in terms of the spherical Bessel function of order zero ($j_0$) namely, 
\begin{equation}\label{bessel}
\rho_\text{FB}(r) = \left\{
\begin{array}{ll}
\displaystyle \sum_{\nu=1}^{n_{max}} a_\nu j_0\left( \frac{\nu\pi r}{R_{cut}} \right), & r\le R_{cut} \\[0.15in]
0, & \text{otherwise},
\end{array}
\right.
\end{equation}
where $a_\nu$ are the expansion coefficients and $R_{cut}$ is a cut-off radius beyond which the charge density is sufficiently small and is equated to zero. The first $n_{max}$ coefficients of this series expansion are obtained directly from the experimental data~\cite{DEVRIES1987495}. Here $n_{max}=R_{cut}q_{max}/\pi$ where $q_{max}$ is the maximum momentum transfer up to which the charge form factor data are determined. For the normalization, the integral of $\rho_\text{FB}(r)$ over all spaces must be equal to the total nuclear charge $+Ze$. 

In the sum-of-Gaussians approach, the charge density is expressed as 
\begin{equation}\label{gaussians}
\rho_\text{SOG}(r) = \sum_{i=1}^{m_{max}} A_i \left\{ \exp\left[-x_i(r)^2\right] + \exp\left[-y_i(r)^2\right] \right\}, 
\end{equation}
where $y_i(r) = (r+R_i)/\gamma$ and $x_i(r)=(r-R_i)/\gamma$. The expansion coefficients $A_i$ are given by 
\begin{equation}
A_i = \frac{ZeQ_i}{2\pi^\frac{3}{2}\gamma^3(1+2R_i^2/\gamma^2)}, 
\end{equation}
where $\gamma$ is the width of the Gaussians and is related to the root-mean-square radius (RMS) through $R_g=\gamma\sqrt{3/2}$. 
According to Sick~\cite{SICK1974509}, $\gamma$ is chosen equal to the smallest width of the peaks in the nuclear radial wave functions calculated using the HF method. 
The author reported that the $\gamma$ values extracted from the harmonic oscillator and WS radial wave functions yield almost identical results. 
The charge fraction $Q_i$ must be normalized such that $\sum_i Q_i = 1$. 

In practice, only first few terms are included within the sum in Eq.~\eqref{bessel} and Eq.~\eqref{gaussians}. 
For example, a truncation with $n_{max}\le 17$ and $m_{max}\le 12$ was imposed for the analyses carried out in Ref.~\cite{DEVRIES1987495}. 
The parameters of $\rho_\text{FB}(r)$ and $\rho_\text{SOG}(r)$ deduced from experiments are given in the data compilation~\cite{DEVRIES1987495}. 

\begin{figure*}[ht!]
\begin{center}
    \includegraphics{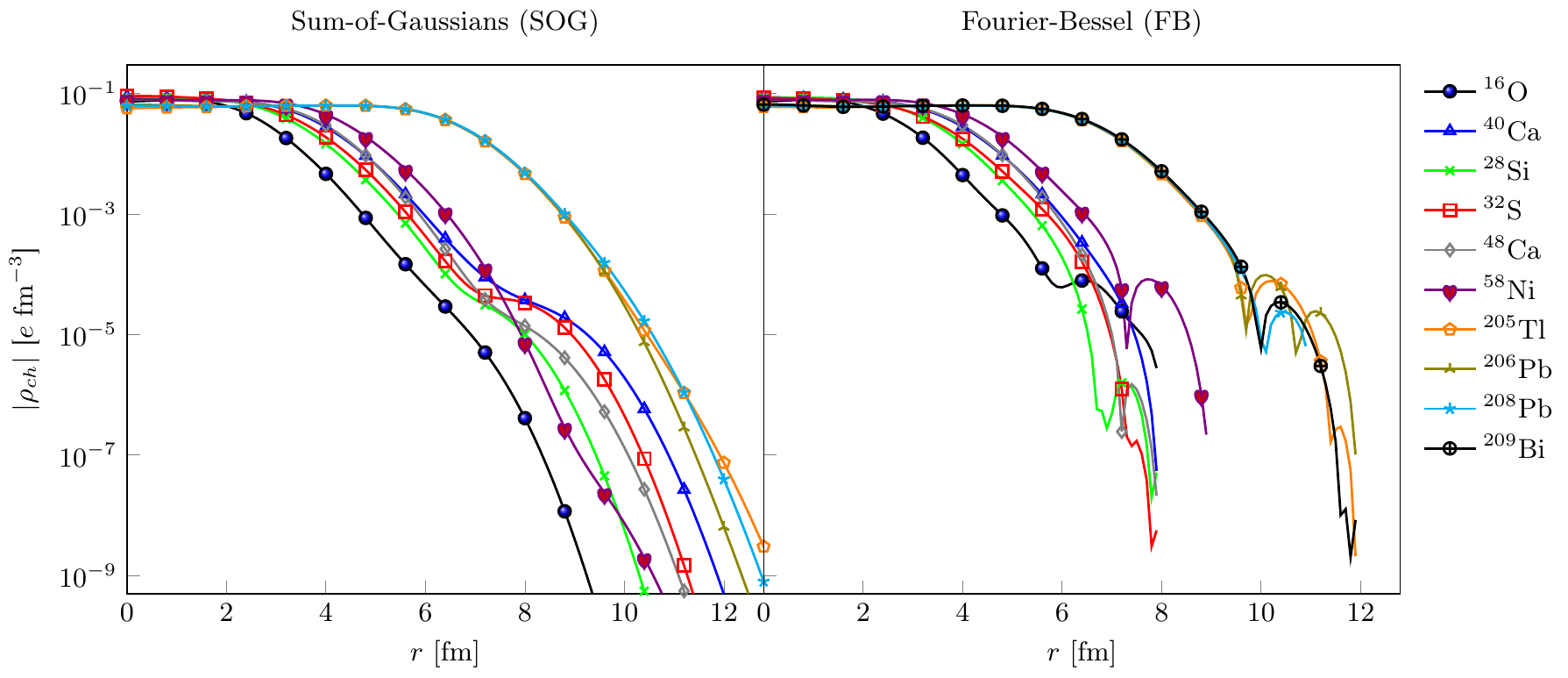}
    \caption{
\label{fig1}(Color online) Illustration of oscillations in the charge densities deduced from electron scattering experiments using the FB (right) and SOG (left) analyses~\cite{DEVRIES1987495}. %The vertical axis represents the absolute charge density values. 
%\tcg{Here I added the y axis label as suggested, could you please check ?}
}
\end{center}
\end{figure*}
 
We remark that the charge density obtained from the FB approach has an undesired property--it contains an oscillatory component and, sometimes, has a negative value near the cut-off radius $R_{cut}$ as illustrated in the right panel of Fig.~\ref{fig1}. 
Although these oscillations are extremely small in magnitude, they are greatly amplified by the first and second derivatives, then leading to a large error, if the GGA is employed for treating the Coulomb exchange term. 
Oscillations in the asymptotics of charge density obtained from the SOG approach are also visible for some nuclei, however their magnitude are generally much smaller (larger width) as shown in the left panel of Fig.~\ref{fig1}. 
This unphysical property originates from the incompleteness in the expansions and too small cut-off radius in the FB analysis. 
Fundamentally, the tail of charge density should decrease monotonically 
%%\tcb{ is monotonically correct? or `exponentially` correct?} 
%%
%%\tcg{As I understand, a function $f$ is said to be monotonically decreasing if $f(r_2)<f(r_1)$ and $r_2>r_1$ for any $r_2$ and $r_1$ in the domain of interest. 
%%So it is generally correct to use `monotonically' here. Fundamentally, the tail of the charge %%density is not an exponential, in general it is a sum of Gaussians (sum of radial wave functions %%squared).} -> Thanks for the answer!!
since the asymptotic radial wave functions decay exponentially. 

In order to avoid this problem, we replace both FB and SOG data from a point just before the oscillations occur (denoted as $R_x$) 
towards infinity with the 2pF function (see subsection~\ref{2pF} for the definition of 2pF). 
We select for both data sets $R_x=4.5$\,fm for $A\le 58$ and $8.5$\,fm for heavier nuclei. Instead of performing a global fit, the 2pF function parameters are here determined by matching 
it as well as its first and second derivatives with those of the data at $R_x$. 
Therefore, the resulting 2pF function might not be optimal for the whole available data in the domain of $[R_x,+\infty]$. 
Nevertheless, it is the only way to ensure the continuity of the charge density at $R_x$. 
This replacement would not cause a serious error since the charge density in this domain is small. 
%\tcb{Question again : your matching guarantees the total number of charge? or Did you renormalize to get the correct number of charge? } 
%\tcg{Yes I did, I normalized them after the modification} ; Thanks !!

\subsection{Uniform charge distribution}\label{ax}

For a sphere of radius $R_C$ containing a total charge of $+Ze$ uniformly distributed throughout its volume, the charge density is written as 
\begin{equation}\label{unif}
\rho_\text{Unif}(r) = \left\{
\begin{array}{ll}
\displaystyle \rho_0, & r\le R_C \\[0.1in]
0, & \text{otherwise}, 
\end{array}
\right.
\end{equation}
and the normalization condition implies that 
\begin{equation}
\rho_0=\frac{3Ze}{4\pi R_C^3}. 
\end{equation}
Here we follow Ref.~\cite{ToHa2002} wherein $\rho_\text{Unif}(r)$ is regarded as a pointed proton distribution and is normalized to $Z$ instead of $+Ze$. Therefore, its RMS radius (denoted as $R_\text{Unif}$) can be calculated analytically, namely 
\begin{equation}\label{rc}
\displaystyle R_\text{Unif} = R_C\sqrt{\frac{3}{5}}.
\end{equation}
In this uniform distribution, the corrections for the finite size, Darwin-Foldy term and COM motion must be introduced in order to convert $R_\text{Unif}$ into the charge radius. 
Eq.~\eqref{rc} provides an experimental constraint for $R_C$, at least, for the cases where the experimental data are available. Otherwise it can be parameterized as usual, namely $R_C=r_C(A-1)^\frac{1}{3}$ where $r_C\approx 1.26$~fm~\cite{SWV}. 
The validity of different methods for determining $R_C$ can also be tested within the framework of the present study. More discussions on this point are given in the next section. 

Because of its simplicity, the uniform charge distribution is widely used in nuclear physics, especially within the nuclear optical model. 
However, one should note that it is an assumption of classical electromagnetism and has no quantum mechanical equivalent since wave functions, which are the building-blocks of charge density, are required to be continuous in coordinate space. 
Furthermore, the two aforementioned approaches to the Coulomb exchange term\,(Slater and GGA) are evidently inapplicable for this distribution because of its discontinuity and nondifferentiability at the surface of a nucleus. 

\subsection{Two-parameter Fermi function}\label{2pF}

We also consider a realistic phenomenological model for the charge density distribution namely the two-parameter Fermi function, 
\begin{equation}\label{fermi}
\displaystyle \rho_{2pF}(r) = \frac{\tilde{\rho}_0}{1+\exp(\frac{r-c}{z})}. 
\end{equation}

Unlike the previous models, the 2pF is continuous and differentiable, and moreover it decays monotonically towards large distances. 
Although the 2pF function looks similar to 
the $f_i(r)$ functions of the WS potential in Eq.~\eqref{fi}, 
%%
%%\tcg{Here I modified the sentence to avoid the term `form factor'.},-> thanks great!!
%%
in general $c$ and $z$ are smaller than the length and the surface diffuseness parameter of the WS potential. 
It is also evident from the Skyrme HF theory that the mean field potential is not a linear function of densities. Thus it is not necessary that their geometrical characterizing parameters a the same with the WS potential parameters. 
More details on this point are given in Ref.~\cite{DG}. 

As its name indicates, the 2pF is determined by two parameters because $\tilde{\rho}_0$ is obtained via the normalization condition, 
\begin{equation}
\tilde{\rho}_0 = \frac{Ze}{4\pi z^3F_2(c/z)}, 
\end{equation}
where $F_2(c/z)$ is the second order Fermi integral (see Appendix C of Ref.~\cite{Elton}). 
Similar to the previous subsection, we employ the convention that $\rho_{2pF}(r)$ is normalized to $Z$, therefore the parameter $c$ can be extracted from the charge radius by solving the following equation~\cite{Elton} 
\begin{equation}
R_{ch}^2 = \frac{4\pi\tilde{\rho}_0 z^5}{Z}F_4\left(\frac{c}{z}\right) + \frac{3}{2}\sum_{i=1}^3 \theta_i r_i^2 + \frac{3}{4}\left( \frac{\hbar}{mc} \right)^2 - \frac{3}{2} \frac{b^2}{A}. 
\end{equation}
As before, the last three terms account for the finite size of proton, Darwin-Foldy term and center-of-mass motion, respectively. 
$F_4(c/z)$ is the fourth order Fermi integral (see Appendix C of Ref.~\cite{Elton}). 

Recently, Horiuchi~\cite{10.1093/ptep/ptab136} proposed a new method for determining the surface thickness of charge density. Implementing the Taylor expansion of $\rho_{2pF}(r)$ at $r=c$ and retaining up to the first-order term, he derived the following relation, 
\begin{equation}\label{diffuseness}
z = -\frac{ \tilde{\rho}_0 }{ 4 } [ \rho_{2pF}'(c) ]^{-1}, 
\end{equation}
where $\rho_{2pF}'(c)$ denotes the first derivative of $\rho_{2pF}(r)$ at $r=c$. By matching $\rho_{2pF}$ on the right-hand-side of Eq.~\eqref{diffuseness} with the charge density constructed from eigenfunctions of the WS potential whose Coulomb term is, in turn, a function of $\rho_{2pF}$, Eq.~\eqref{diffuseness} can be solved in a self-consistent manner. 
This offers an alternative means for constraining $z$ when experimental data are unavailable. 
%For the present study, we use the values of $c$ and $z$ obtained by fitting to the electron scattering data as listed in Table I of Ref.~\cite{DeVries1978}. 

Although this distribution is nicely representative for the diffuseness at the nuclear surface region, it is not able to describe the oscillations of the charge density observed in the nucleus interior. It has been shown that this difficulty can be overcome by extending Eq.~\eqref{fermi} to the three-parameter Fermi (3pF) function~\cite{PhysRevC.90.067304}. 
However, the 3pF is not a good choice for a general application because of lack of experimental data for constraining the third parameter. 

\subsection{Hartree-Fock calculations}

Within a microscopic nuclear structure model, the charge density is basically decomposed into three components 
\begin{equation}
\rho_{ch}(r) = \rho_{ch}^p(r) + \rho_{ch}^n(r) + \rho_{ch}^{ls}(r)
\end{equation}
where $\rho_{ch}^p(r)$/$\rho_{ch}^n(r)$ come from the finite charge distribution of the proton/neutron folded with the point-like proton/neutron density, and $\rho_{ch}^{ls}(r)$ is 
the relativistic electromagnetic correction which depends on the spin-orbit coupling. 
However, the shape of the charge-density distribution is mainly determined by the shape of the point-like proton density. 
The contributions $\rho_{ch}^n(r)$ and $\rho_{ch}^{ls}(r)$ were found to be negligible~\cite{XaNa2022}. 
Furthermore, they tend to cancel out each other, thus will not be considered here. 

By definition $\rho_{ch}^p(r)$ is given by 
\begin{equation}\label{chp}
\rho_{ch}^p(r) = \int d\boldsymbol{r}' \rho_p(r')G_p(\boldsymbol{r}-\boldsymbol{r}'), 
\end{equation}
where $\boldsymbol{r}$ is the position vector in $\mathbb{R}^3$. The effective electromagnetic form factor $G_p$ is taken as a sum of three Gaussians as described in Ref.~\cite{BABrown_1979}. 
The point-like proton density $\rho_p(r)$ can be defined in terms of proton radial wave functions $R_\alpha(r)$ namely
\begin{equation}
\rho_p(r) = \frac{1}{4\pi} \sum_\alpha n_\alpha |R_\alpha(r)|^2, 
\end{equation}
where $\alpha$ stands for the spherical quantum numbers $nlj$ and the sum runs over all occupied states. 
The proton occupation number $n_\alpha$ is obtained with the so-called equal-filling approximation. Therefore, for a closed-shell configuration, $n_\alpha=(2j+1)$ for the occupied orbits and 0 for the unoccupied orbits. 
%An alternative method for determining $n_\alpha$ is discussed in Ref.~\cite{PhysRevC.67.034317}.
%\tcb{`An alternative method for determining', <- Is this sentence really necessary?}

The radial wave functions $R_\alpha(r)$ are taken as the eigenfunctions of the Skyrme HF mean field. 
For the Skyrme Hamiltonian, we select the SLY5 parameter set~\cite{CHABANAT1998231} which is invariant under rotation in the isospin space, while the Coulomb exchange term is treated using the Slater approximation. 
The charge-symmetry and charge-independence breaking forces~\cite{SAGAWA19957} are neglected. We use the HFBRAD program~\cite{BENNACEUR200596} to solve the spherical Skyrme HF equation with which the Coulomb terms are evaluated using the point-like proton density. The finite size effect is corrected for by using Eq.~\eqref{chp} within an external program after the HF variation is terminated. 

\begin{figure*}[ht!]
\begin{center}
    \includegraphics[scale=0.77]{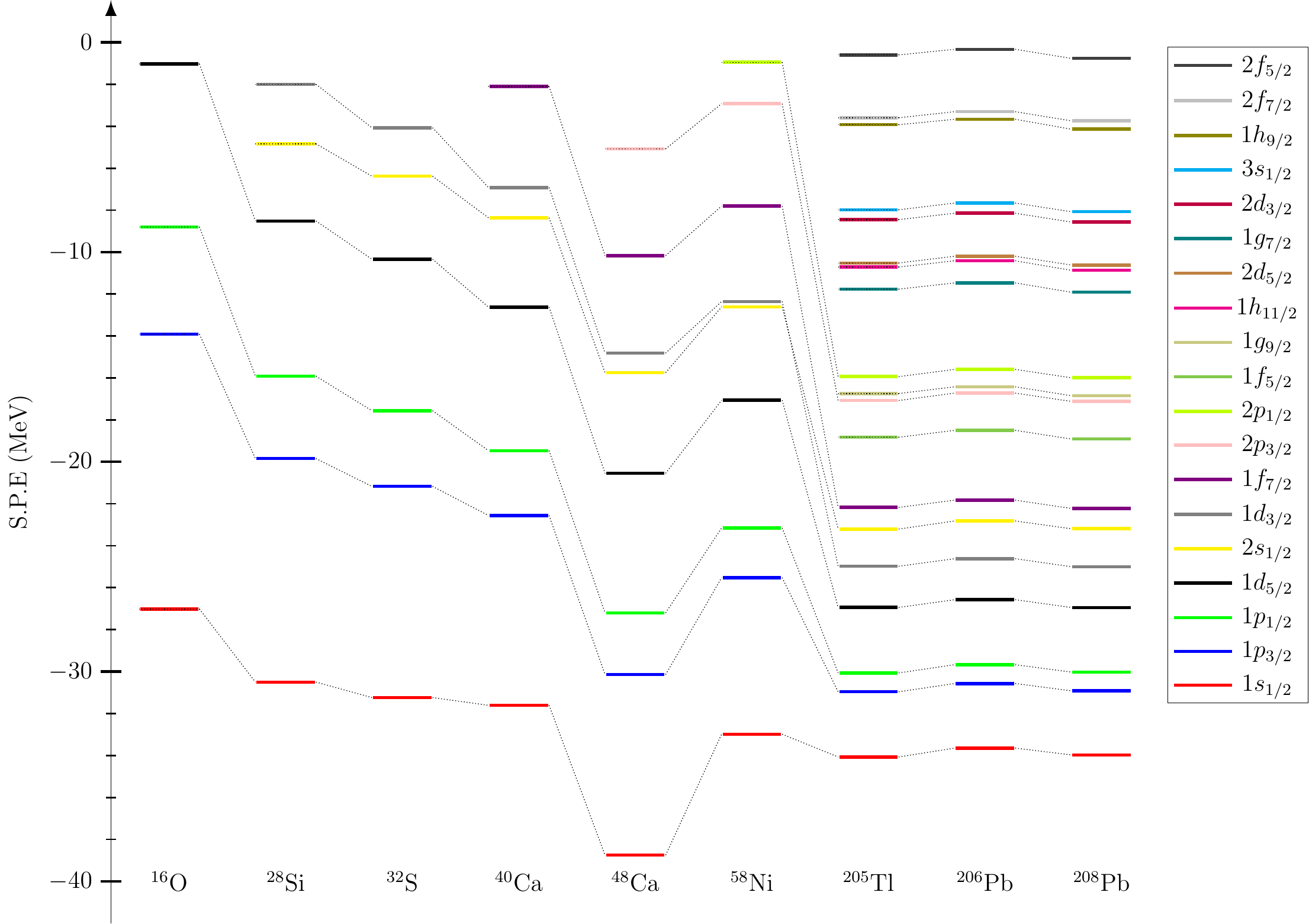}
    \caption{
\label{fig2}(Color online) Averaged proton single-particle energies between the refined FB and SOG data for  charge density. For this calculation, the Coulomb exchange term is evaluated with the GGA and the spin-orbit interaction is included. }
\end{center}
\end{figure*}

\section{Comparative study of Coulomb potential}\label{sec:coulombpot}

This section starts with an inspection of the uncertainties on the experimental data for charge density as well as their impact on proton single-particle energies. As a next step, the validity of the charge density models listed in section~\ref{input} as an input for Coulomb potential will be checked, through the comparison of their prediction for proton energy levels to that of the model-independent data. Likewise, the approximate Coulomb functionals including that of Eq.~\eqref{eq1} and those reviewed in section~\ref{func} will be tested by comparing their prediction with the result obtained using an exact treatment of the Coulomb exchange term. 
%
%\tcg{To be clearer, some extensions were introduced to this paragraph. Could you please take a look ?}; Yes, it is good, thanks

\subsection{Inspection of experimental charge density data}

Before performing a comparative study of Coulomb potential, it is necessary to figure out the impact of the differences between the FB and SOG data for charge density on proton single-particle energies.
%%\tcb{I'm confused with the `observable'. So the observable is the physical quantities which we can measure. 
%%So you mentioned about RMS or neutron skin. But the above sentences 'on a selected observable', 
%%I thought that you are talking about some nucleus which we can check with. 
%%It might be better to use `as a selected observable' instead of `on a selected observable'.} %%\tcg{Thank you so much for this comment ! I have replaced `a selected observable' with `proton %%single-particle energies'.}
%%
For simplicity, we neglect the uncertainties on the coefficients of FB and SOG expansions even though they are available from the data compilations. Noted that the quantity which will be considered throughout this section is the proton single-particle energies. The other observables such as the RMS radii and neutron skin thicknesses were found to be much less sensitive to small variation in the Coulomb terms. 

Furthermore, instead of a point-by-point comparison, we define the following radial mismatch factor, 
\begin{equation}
\Lambda = \frac{ \bar{\Omega} - \Omega }{ \bar{\Omega} }, 
\end{equation}
 as an effective measure for the charge density differences, 
where $\Omega$ is the overlap integral between the FB and SOG data:  
\begin{equation}
\Omega = 4\pi\int_0^\infty \rho_\text{FB}(r) \rho_\text{SOG}(r) r^2 dr, 
\end{equation}
whereas $\bar{\Omega}$ is the integral of their average squared: 
\begin{equation}
\displaystyle
\bar{\Omega} = 4\pi\int_0^\infty \left[ \frac{\rho_\text{FB}(r) + \rho_\text{SOG}(r)}{2} \right]^2 r^2 dr. 
\end{equation} 
The resulting $\Lambda$ values in \% are 3.441$\times 10^{-3}$ ($^{16}$O), 5.659$\times 10^{-3}$ ($^{28}$Si), 0.209 ($^{32}$S), 2.601$\times 10^{-2}$ ($^{40}$Ca), 4.126$\times 10^{-2}$ ($^{48}$Ca), 3.866$\times 10^{-3}$ ($^{58}$Ni), 3.641$\times 10^{-4}$ ($^{205}$Tl), 5.048$\times 10^{-4}$ ($^{206}$Pb) and 1.459$\times 10^{-2}$ ($^{208}$Pb). 
%\tcb{I think this is a really good comparison. Q: is this what you invented ? or some %conventional things to use. I'm also wondering the unit is correct. Probably correct when we %think the total number of charge. What about using $ \int (\rho_{\rm FB} -\rho_{\rm SOG})^2 r^2\,dr $/$ \int \rho_{\rm FB}^2 r^2\,dr $?}
%\tcg{Before, we used this redial mismatch factor $\Lambda$ as an effective measure of the %differences between proton and neutron radial wave functions which are induced by the isospin-%symmetry breaking. I think it is used here for first time for the charge density comparison. %About the unit, $\Lambda$ is the relative deviation between $\bar{\Omega}$ and $\Omega$ so it has %no unit ($\bar{\Omega}$ and $\Omega$ have the same unit). For your suggested formula $ \int (\rho_{\rm FB} -\rho_{\rm SOG})^2 r^2\,dr $/$ \int \rho_{\rm FB}^2 r^2\,dr $, I think it will give %slightly different values compared with $\Lambda$, and probably it will also provide a good %measure for the charge density differences. Your formula has been used in Ref. %https://doi.org/10.1103/PhysRevC.99.024309 to measure the differences between the GGA and Slater %densities, except that the integrals are omitted.}
%%

We recall that the exact Coulomb exchange term is nonlocal in coordinate spaces and cannot be written as a function of charge density, it is thus inappropriate for the present study which replaces the self-consistent HF mean field with the phenomenological WS potential. In order to demonstrate whether or not the selection of Coulomb exchange functional is matter to the impact of the FB-SOG charge density differences on proton single-particle energies, 
%\tcb{I'm not quite sure about the above sentence. -> 
%In order to demonstrate the variation in Coulomb exchange term 
%causes the differences on the proton-single particle energies when the FB or SOG charge density
%is utilized?}
%%
%\tcg{No, your suggested sentence here has a different meaning. To be clear, I have made a modification in red. Could you please check it ?} Ok,,, thanks
%%
%%
we consider three different approximations for Coulomb potential: one of them consists purely of the direct term whereas the other two include also an exchange term employing either the Slater approximation or the GGA. Then together with the nuclear component described in Eq.~\eqref{ws}, the single-particle energies and wave functions can be obtained by solving the radial Schr\"{o}dinger's equation. The splits in proton energies due to the differences in the charge density data are listed in Table~\ref{tab0}. As a convention, the negative sign indicates that the FB data yields a lower proton energy level relative to that yielded by the SOG data i.e., $\Delta E^{\rm FB}_{\rm SOG} = E_{\rm FB} - E_{\rm SOG}$. The meaning of the positive sign is opposite. For completeness, the averaged proton energies between the FB and SOG results calculated using the GGA functional 
[denoted as $\bar{E}^{\rm FB}_{\rm SOG} = (E_{\rm FB} + E_{\rm SOG})/2$]
%%
%%\tcg{Here I added an Eq. Please check ?} -> Thanks, great!
%%
are graphically illustrated in Fig.~\ref{fig2}. It is clearly observed from Table~\ref{tab0} that the splits in proton energies induced by the charge density differences are insensitive to the Coulomb exchange term. This witnesses that any existing Coulomb functionals can be employed for our test of the charge density models which will be discussed in the following subsection. 

We remark that although the amount of energy splits is mainly determined by the size of $\Lambda$, it can be greatly enhanced by the shell-structure effect. As a general feature, the solution of the radial Schr\"{o}dinger equation for protons will be particularly sensitive to the Coulomb terms when the last occupied orbit is highly filled and at the same time, has a low centrifugal barrier (low orbital angular momentum). 
In the remainder of this paper, this phenomenon will be referred to as the ``weakly bound effect''. 
%%
%%\tcg{Here I added a sentence describing the `weakly bound effect'. Could you please check ? } 
%% Thanks, great !!
%%
This is the main reason behind the large energy split of about $-500$~keV for $^{32}$S. 
We notice that this number is comparable to the energy splits due to the use of the Slater approximation within the Skyrme HF framework studied in Ref.~\cite{PhysRevC.63.024312}. Therefore, a special attention must be paid to the inclusion of this nucleus for our analyses in the following subsection. The impact for the two Calcium\,($^{40}\rm{Ca}$,$^{48}\rm{Ca}$) isotopes is also quite large namely about $-100$~keV, while the impact for the other cases is less than 50~keV in magnitude.  

\subsection{Test of charge density models}\label{test1}

Because of its discontinuity and nondifferentiability, the uniform distribution is inappropriate to be used as an input for calculating the Coulomb exchange term within any existing approximations. In order to evaluate the Coulomb potential for this hypothetical distribution as well as the other charge density models listed in section~\ref{input} on an equal footing, the Coulomb exchange term will be omitted for our study in this subsection. It was shown in the previous subsection that the absence of the Coulomb exchange term does not significantly affect the energy splits induced by the differences in the charge density data\,($\Delta E^{\rm FB}_{\rm SOG}$), even for $^{32}$S where $\Lambda$ is as large as 0.209~\%. 

Here we use the averaged proton energy between FB and SOG values as a reference\,(denoted as $\bar{E}_R$) relative to which the result obtained for a given charge density model is compared. 
As an exception, only the FB data are used for $^{209}$Bi because a SOG analysis has not been carried out for this nucleus. 
The proton energies evaluated with the uniform distribution, the 2pF function and the microscopic Skyrme-HF charge density are denoted as $E_{\rm Unif}$, $E_{2pF}$ and $E_{HF}$\footnote{$E_{HF}$ denotes the proton single-particle energies evaluated with the WS potential as a nuclear component and the charge density obtained from the microscopic Skyrme HF calculation as an input for the Coulomb potential. It should not be confused with the eigenvalues of the HF mean field.}, respectively. 
Our results for the proton energy differences including $\Delta E^{\rm Unif}_{R}=E_{\rm Unif}-\bar{E}_R$, $\Delta E^{2pF}_{R}=E_{2pF}-\bar{E}_R$ and $\Delta E^{HF}_{R}=E_{HF}-\bar{E}_R$ are given in Table~\ref{tab1}. 
%\tcg{``As a sign convention, a negative value indicates that the charge density model yields a lower proton energy level relative to that yielded by the experimental data. The opposite is true for a positive value''}. 
%\tcg{Here I made a considerable modification. Could you please check ?}-> YH; I like this
%%

We remark that the energy differences obtained in this subsection for all charge-density models are generally small in magnitude. 
For most cases, their magnitude is only of a few tens keV larger than those induced by the experimental uncertainties obtained in the previous subsection\,($\Delta E^{\rm FB}_{\rm SOG}$). 
The values obtained for $^{32}$S are somewhat larger compared with the other nuclei and positive for all bound orbitals, because of large uncertainties in the model-independent data for charge density and an additional enhancement due to the weakly bound effect. 
%\tcb{How do you define the energy difference? I want to make it clear. 
%What is the weakly bound effect in this case?} 
%%
%\tcg{I added a description of ``'weakly bound effect' in the last paragraph of the previous subsection, could you please check it ?} YH; thanks I've checked it.
%%
It is also seen that all the selected charge density models provide a similar accuracy for proton single-particle energies. 
In particular, the microscopic Skyrme HF model for charge density works adequately regardless of the various deficiencies related to the isospin-symmetry breaking discussed in Ref.\,\cite{XaNa2022}. 
We notice that, in general the calculations using the uniform distribution or the 2pF function are strongly parameter-dependent. 
For example, if the Coulomb radius is parameterized as $R_C\approx 1.26~\text{fm} \times (A-1)^\frac{1}{3}$ as usual, considerably larger proton single-particle energy differences are obtained in many cases. 
Similar problem was also observed within the 2pF function. 
Therefore, it is necessary to constrain their parameters using the relevant experimental data if these distribution functions are selected as a charge-density model. 

\subsection{Test of Coulomb functionals}\label{test2}

The present subsection is concerned with the compensation for the omission of Coulomb exchange term 
which is unavoidable when using the uniform distribution for charge densities. We are interested specifically in the method employed in Eq.~\eqref{eq1} by excluding the last proton contribution to Coulomb direct term. In order to verify this traditional idea of self-interaction correction, it is instructive to perform an analytic analysis of its impact before discussing the numerical results. 

Since the Coulomb direct term~\eqref{dir} is linear in charge density, it can be written as 
\begin{equation}
V_{dir}[\rho_{ch}^Z] = V_{dir}[\rho_{ch}^{Z-1}] + V_{dir}[\rho_{ch}^1]
\end{equation}
where $\rho_{ch}^Z$ is the total charge density which can be decomposed as 
\begin{equation}
\rho_{ch}^Z = \rho_{ch}^{Z-1} + \rho_{ch}^1
\end{equation}
with $\rho_{ch}^1$ being the last proton contribution and $\rho_{ch}^{Z-1}$ the contribution of the remaining $Z-1$ protons. In order to gain a better insight into the functional structure of the Coulomb direct term, we further assume that $\rho_{ch}^Z$ and $\rho_{ch}^{Z-1}$ have identical radial form and they differ from each other only by the normalization condition, 
%% \tcg{Here I replaced `form factor' with `form', Could you please check ?} ;YH-> is functional form appropriate?
namely $\rho_{ch}^Z$ is normalized to $Z$ whereas $\rho_{ch}^{Z-1}$ is normalized to $Z-1$ (this is equivalent to the assumption that each proton has an equal contribution to the total charge density). Subsequently, the following relations can be derived
\begin{equation}
\rho_{ch}^{Z-1} = \left( \frac{Z-1}{Z} \right) \rho_{ch}^Z = (Z-1)\rho_{ch}^1. 
\end{equation}
Because of the linearity, the corresponding Coulomb direct terms can be evaluated as
\begin{equation}
V_{dir}[\rho_{ch}^{Z-1}] = \left( \frac{Z-1}{Z}\right) V_{dir}[\rho_{ch}^Z]=(Z-1)V_{dir}[\rho_{ch}^1]. 
\end{equation}

Based on this assumption, the self-interaction correction discussed above can be done simply by replacing $V_{dir}[\rho_{ch}^Z]$ with $V_{dir}[\rho_{ch}^Z]\times(Z-1)/Z$. Note that the Coulomb potential in Eq.~\eqref{eq1} fully satisfies these properties with $\rho_{ch}^Z$ being the uniform distribution. 
%%
%%\tcg{To be clearer, I made an extension to the sentence here. Could you please check ?}; YH;-> ok now..
%%
It is interesting to remark that the factor $(Z-1)/Z$ goes to 1 as $Z$ goes to infinity, indicating that this correction has a greatest effect in light nuclei. We found that this replacement leads to lower proton energy levels, namely by about 500~keV in $^{16}$O and 200~keV in $^{208}$Pb, relative to the levels obtained with the pure Coulomb direct term $V_{dir}[\rho_{ch}^Z]$. 

Throughout this subsection, the Coulomb functional obtained in this way will be referred to as ``traditional functional''. In order to make a numerical test this traditional approach, it is useful to decompose the proton energy difference for a given orbital as
\begin{equation}\label{diff}
\Delta E^T_F=\Delta E^T_G + \Delta E^G_S + \Delta E^S_F, 
\end{equation}
where $\Delta E^T_F$ is the single-particle energy difference between the traditional functional and the exact Fock term. 
On the right-hand-side, $\Delta E^T_G$, $\Delta E^G_S$ and $\Delta E^S_F$ are, respectively, the single-particle energy difference of the traditional treatment relative to GGA, GGA relative to Slater approximation, and Slater relative to the exact Fock term. Explicitly, $\Delta E^T_F = E^T - E^F$, $\Delta E^T_G = E^T - E^G$, $\Delta E^G_S = E^G - E^S$ and $\Delta E^S_F = E^S - E^F$.
As the uniform distribution is inappropriate for the evaluation of Coulomb exchange term, the 2pF function will be selected instead for this test. 

We notice that no exact treatment of Coulomb exchange term is performed in the present work, the data of $\Delta E^S_F$ for $^{16}$O, $^{40}$Ca, $^{48}$Ni and $^{208}$Pb are taken from Ref.~\cite{PhysRevC.63.024312}. Since the mass dependence of $\Delta E^S_F$ is not so strong, it is reasonable to use the values obtained for $^{40}$Ca for $^{28}$Si, $^{32}$S and $^{48}$Ca which were not considered in Ref.~\cite{PhysRevC.63.024312} because they reside in the neighbor in the nuclear chart. 
For the same reason, we use the $\Delta E^S_F$ values obtained for $^{208}$Pb for the remaining nuclei. One may argue here that the proton energy differences may depend on the nuclear component as well as the method for solving the Schr\"{o}dinger equation, or they may vary significantly when transferring from a self-consistent to a phenomenological mean field. We have checked such dependence by looking at the term $\Delta E^G_S$, we found that its values obtained for $^{208}$Pb using WS potential are in the range between -1 and 19~keV, which are in remarkably good agreement with those calculated within the self-consistent Skyrme HF method. 

Our numerical results for $\Delta E^T_F$ as well as for $\Delta E^T_G$ and $\Delta E^G_S$ are given in Table~\ref{tab2}. It is seen that in light nuclei (around $Z=8$) negative $\Delta E^T_G$ values are obtained because the traditional self-interaction correction is stronger than the GGA exchange term. After $Z=8$, $\Delta E^T_G$ increases gradually with the atomic number and finally reaches a saturation at around $Z=80$. The saturated value of $\Delta E^T_G$ is about 300~keV. We notice also that the functional-driven energy differences are insensitive to the weakly bound effect, 
because the values obtained for $^{32}$S do not differ significantly from those of the neighboring nuclei as observed in the two previous subsections. Combining these values with those of $\Delta E^S_F$ from the above-cited self-consistent calculations, we obtained that the total energy differences between the traditional correction and the exact treatment are in the range between $-130$ and 620~keV for light nuclei, and 136 and 800~keV for heavier nuclei. A larger value is expected if those induced by the charge density are added together. Therefore, the expression Eq.~\eqref{eq1} for Coulomb potential should not be applied for a high precision calculation such as the shell-model description of the isospin-symmetry breaking. 

\section{Conclusion}\label{sec:con}

We have performed a comparative study of Coulomb potential for various charge-density models, including the uniform distribution, the 2pF function, and the microscopic Skyrme HF calculation. 
The phenomenological WS potential was selected as a nuclear component of the mean-field Hamiltonian. We found that the differences between the proton single-particle energies produced with these charge-density models and those yielded by the model-independent data are relatively small. 
Generally, these differences are less than 100~keV in magnitude, except for $^{32}$S because of high sensitivity to small variations in the potential occurred when the last occupied state is fully filled and has no centrifugal barrier. 
Although the proton single-particle energies obtained in this work using the uniform distribution and the 2pF function are remarkably accurate, in general this observable is strongly parameter-dependent.
Thus, it is highly recommended to constraint the charge-density model parameters case-by-case using the available experimental data on charge radii as illustrated in the present paper. Otherwise the microscopic Skyrme-HF model should be employed instead. 
%%
%all models works quite well. \tcb{all models works quite well -> the energy differences among different density models are quite small.} 
%%
%\tcg{Sorry, I realized that this paragraph is not clear, so I have made a considerable modification here. Could you please take a look at the above paragraph in red ?} YH; thank for the new paragraph. You dont' need to be sorry in the paper. We all have different point of view and it is always difficult to make people understood.

% On average, the Skyrme HF model yields the best result 
%\tcb{What is the meaing of the `best'? Could you make it clear?}
%%
%regardless of the minor deficiencies relating to the isospin-symmetry breaking as discussed in Ref.~\cite{XaNa2022}. 

In addition, we have also studied the traditional approach to compensate for the omission of the exchange term by excluding the last proton contribution. This approach is commonly employed when the uniform distribution is selected as a charge-density model. 
%%
%%\tcg{To be clearer, I added a new sentence here. Could you please check it ?} YH; thanks
%%
We found that such compensation can be implemented by simply multiplying the Coulomb direct term with the factor $(Z-1)/Z$. Therefore, it can be regarded as a variation in Coulomb functional. Combining our numerical result with that of Ref.~\cite{PhysRevC.63.024312}, the proton energy levels obtained with this traditional functional are underbound by 100 to 800~keV for nuclei with $Z\ge 28$, relative to the those obtained with an exact treatment. The opposite pattern trends to appear in the lighter $Z$ region where the factor $(Z-1)/Z$ reduces significantly from unity, for example the $1d_{5/2}$ level in $^{16}$O obtained with the traditional functional is overbound by 254~keV. It is also seen from our results that the functional-driven energy differences are rather insensitive to the weakly-bound effect since the values obtained for $^{32}$S where this effect is expected to be strongest do not alter significantly from those of its neighboring nuclei. 

As a final conclusion, the use of Eq.~\eqref{eq1} for a precision calculation is at risk. Instead of this traditional formula, it is desired to evaluate the Coulomb potential using realistic charge density models such as the 2pF function or Skyrme HF, as well as more fundamental approaches for Coulomb-exchange functional such as the Slater or the GGA.

\begin{acknowledgments} 

%This work is supported in part by the Ewha Womans University Research Grant of 2021(1-2021-0520-001-1) and
This work is supported by the National Research Foundation of Korea(NRF) grant funded by the Korea government(MSIT)(No. 2021R1A2C2094378).

\end{acknowledgments}

\bibliography{Coulomb}

\appendix*
\section{Numerical results}

\begin{table*}[ht!]
\caption{\label{tab0} Differences\,($\Delta E^{\rm FB}_{\rm SOG} = E_{\rm FB} - E_{\rm SOG}$) in proton single-particle energies induced by the differences between the FB and SOG charge-density data. 
%The negative values indicate that the proton levels obtained with the FB data are lower than those obtained with the SOG data. 
The calculation without Coulomb exchange term is labeled with `Direct', whereas those performed using the GGA and the Slater approximation are labeled with `GGA' and `Slater', respectively. The listed numbers are in keV unit.}
\begin{ruledtabular}
\begin{tabular}{c|ccc|ccc|ccc|ccc|ccc} 
Orbital	&	Direct	&	Slater	&	GGA	&	Direct	&	Slater	&	GGA	&	Direct	&	Slater	&	GGA	&	Direct	&	Slater	&	GGA	&	Direct	&	Slater	&	GGA	\\
\hline
	& \multicolumn{3}{c|}{$^{16}$O}	 & \multicolumn{3}{c|}{$^{28}$Si}	& \multicolumn{3}{c|}{$^{32}$S}					&	\multicolumn{3}{c|}{$^{40}$Ca}	 & \multicolumn{3}{c}{$^{48}$Ca} \\																													
$2p_{3/2}$	&		&		&		&		&		&		&		&		&		&		&		&		&	-78	&	-80	&	-71	\\
$1f_{7/2}$	&		&		&		&		&		&		&		&		&		&	-59	&	-58	&	-53	&	-74	&	-73	&	-66	\\
$1d_{3/2}$	&		&		&		&	-25	&	-26	&	-20	&	-452	&	-452	&	-452	&	-75	&	-72	&	-70	&	-94	&	-91	&	-86	\\
$2s_{1/2}$	&		&		&		&	-28	&	-28	&	-20	&	-462	&	-462	&	-462	&	-82	&	-81	&	-76	&	-102	&	-100	&	-95	\\
$1d_{5/2}$	&	2	&	3	&	0	&	-28	&	-27	&	-23	&	-460	&	-460	&	-460	&	-75	&	-73	&	-70	&	-92	&	-88	&	-84	\\
$1p_{1/2}$	&	4	&	3	&	3	&	-36	&	-34	&	-33	&	-514	&	-514	&	-514	&	-96	&	-91	&	-91	&	-115	&	-110	&	-108	\\
$1p_{3/2}$	&	4	&	4	&	3	&	-36	&	-34	&	-32	&	-511	&	-511	&	-511	&	-94	&	-90	&	-89	&	-112	&	-107	&	-105	\\
$1s_{1/2}$	&	8	&	7	&	6	&	-46	&	-43	&	-42	&	-570	&	-570	&	-570	&	-118	&	-112	&	-111	&	-138	&	-130	&	-130	\\
\hline
	&	\multicolumn{3}{c|}{$^{54}$Ni}	& \multicolumn{3}{c|}{$^{205}$Tl}	&	\multicolumn{3}{c|}{$^{206}$Pb}	&	\multicolumn{3}{c|}{$^{208}$Pb}	& \multicolumn{3}{c}{$^{209}$Bi}	\\																														
$2f_{5/2}$	&		&		&		&	-13	&	-14	&	-14	&	-16	&	-18	&		&	-16	&	-18	&	-19	&		&		&		\\
$2f_{7/2}$	&		&		&		&	-13	&	-14	&	-13	&	-17	&	-19	&	-19	&	-18	&	-19	&	-19	&		&		&		\\
$1h_{9/2}$	&		&		&		&	-12	&	-13	&	-12	&	-16	&	-17	&	-18	&	-16	&	-17	&	-17	&		&		&		\\
$3s_{1/2}$	&		&		&		&	-17	&	-17	&	-18	&	-21	&	-22	&	-23	&	-21	&	-22	&	-22	&		&		&		\\
$2d_{3/2}$	&		&		&		&	-16	&	-16	&	-16	&	-20	&	-21	&	-21	&	-20	&	-21	&	-21	&		&		&		\\
$1g_{7/2}$	&		&		&		&	-13	&	-14	&	-13	&	-18	&	-19	&	-19	&	-18	&	-18	&	-19	&		&		&		\\
$2d_{5/2}$	&		&		&		&	-16	&	-15	&	-16	&	-20	&	-21	&	-21	&	-20	&	-21	&	-21	&		&		&		\\
$1h_{11/2}$	&		&		&		&	-11	&	-11	&	-12	&	-15	&	-16	&	-17	&	-16	&	-16	&	-16	&		&		&		\\
$1g_{9/2}$	&		&		&		&	-12	&	-12	&	-13	&	-17	&	-17	&	-18	&	-17	&	-18	&	-18	&		&		&		\\
$1f_{5/2}$	&		&		&		&	-14	&	-14	&	-15	&	-20	&	-20	&	-21	&	-20	&	-20	&	-21	&		&		&		\\
$2p_{1/2}$	&	29	&	30	&	27	&	-18	&	-18	&	-19	&	-23	&	-23	&	-24	&	-22	&	-23	&	-24	&		&		&		\\
$2p_{3/2}$	&	31	&	32	&	29	&	-18	&	-18	&	-18	&	-22	&	-24	&	-24	&	-23	&	-24	&	-23	&		&		&		\\
$1f_{7/2}$	&	30	&	29	&	27	&	-13	&	-14	&	-14	&	-19	&	-19	&	-27	&	-23	&	-19	&	-19	&		&		&		\\
$1d_{3/2}$	&	37	&	35	&	34	&	-16	&	-16	&	-16	&	-22	&	-22	&	-23	&	-22	&	-22	&	-22	&		&		&		\\
$2s_{1/2}$	&	40	&	39	&	38	&	-20	&	-20	&	-20	&	-25	&	-26	&	-27	&	-25	&	-25	&	-27	&		&		&		\\
$1d_{5/2}$	&	36	&	34	&	34	&	-16	&	-15	&	-16	&	-21	&	-21	&	-22	&	-21	&	-21	&	-21	&		&		&		\\
$1p_{1/2}$	&	44	&	42	&	42	&	-17	&	-18	&	-18	&	-24	&	-25	&	-25	&	-23	&	-24	&	-24	&		&		&		\\
$1p_{3/2}$	&	43	&	41	&	41	&	-18	&	-17	&	-17	&	-23	&	-24	&	-24	&	-23	&	-23	&	-24	&		&		&		\\
$1s_{1/2}$	&	52	&	50	&	49	&	-20	&	-20	&	-20	&	-26	&	-27	&	-27	&	-27	&	-26	&	-27	&		&		&		\\
\end{tabular}
\end{ruledtabular}
\end{table*}																													
\begin{table*}[ht!]
\caption{\label{tab1} Proton single-particle energy differences induced by the deviations of the charge-density models relative to the model-independent data. Where $\Delta E^{\rm Unif}_R$, $\Delta E^{2pF}_R$ and $\Delta E^{HF}_R$ correspond to the uniform distribution, the 2pF function and the Skyrme-HF calculation of charge density, respectively (see subsection~\ref{test1} for detailed description). The Coulomb exchange potential is excluded for these calculations. The listed numbers are in keV unit.}
\begin{ruledtabular}
\begin{tabular}{c|ccc|ccc|ccc|ccc|ccc} 
%\backslashbox{Orbital}{Method}
Orbital	&	$\Delta E^{\rm Unif}_R$	&	$\Delta E^{2pF}_R$ & $\Delta E^{HF}_R$	&	$\Delta E^{\rm Unif}_R$	&	$\Delta E^{2pF}_R$ & $\Delta E^{HF}_R$	&	$\Delta E^{\rm Unif}_R$	&	$\Delta E^{2pF}_R$ & $\Delta E^{HF}_R$	&	$\Delta E^{\rm Unif}_R$	&	$\Delta E^{2pF}_R$ & $\Delta E^{HF}_R$	&	$\Delta E^{\rm Unif}_R$	&	$\Delta E^{2pF}_R$ & $\Delta E^{HF}_R$	\\
\hline
	& \multicolumn{3}{c|}{$^{16}$O}	 & \multicolumn{3}{c|}{$^{28}$Si}	& \multicolumn{3}{c|}{$^{32}$S}					&	\multicolumn{3}{c|}{$^{40}$Ca}	 & \multicolumn{3}{c}{$^{48}$Ca} \\
$2p_{3/2}$	&		&		&		&		&		&		&		&		&		&		&		&		&	-4	&	7	&	11	\\
$1f_{7/2}$	&		&		&		&		&		&		&		&		&		&	73	&	6	&	46	&	113	&	13	&	13	\\
$1d_{3/2}$	&		&		&		&	29	&	-18	&	23	&	667	&	631	&	693	&	35	&	8	&	47	&	85	&	24	&	5	\\
$2s_{1/2}$	&		&		&		&	-69	&	-34	&	3	&	540	&	599	&	694	&	-72	&	-17	&	48	&	-6	&	19	&	8	\\
$1d_{5/2}$	&	-1	&	-22	&	-10	&	33	&	-16	&	24	&	689	&	651	&	712	&	42	&	8	&	47	&	93	&	22	&	5	\\
$1p_{1/2}$	&	-40	&	-21	&	-30	&	-38	&	-22	&	13	&	682	&	726	&	804	&	-41	&	3	&	47	&	34	&	34	&	-5	\\
$1p_{3/2}$	&	-42	&	-20	&	-32	&	-32	&	-21	&	15	&	689	&	725	&	801	&	-31	&	4	&	47	&	43	&	32	&	-3	\\
$1s_{1/2}$	&	-144	&	-17	&	-60	&	-159	&	-37	&	-8	&	618	&	783	&	894	&	-158	&	-12	&	48	&	-50	&	40	&	-10	\\
\hline
	&	\multicolumn{3}{c|}{$^{54}$Ni}	& \multicolumn{3}{c|}{$^{205}$Tl}	&	\multicolumn{3}{c|}{$^{206}$Pb}	&	\multicolumn{3}{c|}{$^{208}$Pb}	& \multicolumn{3}{c}{$^{209}$Bi}	\\																														
$2f_{5/2}$	&		&		&		&	-70	&	-48	&	-15	&		&		&		&	-37	&	-72	&	-9	&	-55	&	-42	&	20	\\
$2f_{7/2}$	&		&		&		&	-78	&	-50	&	-18	&	-62	&	-46	&	2	&	-46	&	-74	&	-11	&	-64	&	-44	&	18	\\
$1h_{9/2}$	&		&		&		&	41	&	-46	&	-8	&	63	&	-39	&	3	&	84	&	-59	&	-9	&	58	&	-38	&	15	\\
$3s_{1/2}$	&		&		&		&	-122	&	-31	&	-33	&	-120	&	-40	&	5	&	-109	&	-74	&	-9	&	-116	&	-33	&	30	\\
$2d_{3/2}$	&		&		&		&	-83	&	-42	&	-23	&	-69	&	-42	&	1	&	-56	&	-72	&	-12	&	-70	&	-38	&	20	\\
$1g_{7/2}$	&		&		&		&	20	&	-47	&	-15	&	41	&	-39	&	-2	&	59	&	-63	&	-15	&	35	&	-40	&	10	\\
$2d_{5/2}$	&		&		&		&	-91	&	-44	&	-25	&	-78	&	-41	&	-1	&	-63	&	-73	&	-13	&	-78	&	-39	&	20	\\
$1h_{11/2}$	&		&		&		&	59	&	-45	&	-2	&	81	&	-38	&	8	&	105	&	-56	&	-3	&	76	&	-38	&	18	\\
$1g_{9/2}$	&		&		&		&	41	&	-46	&	-9	&	63	&	-39	&	3	&	84	&	-60	&	-10	&	57	&	-39	&	14	\\
$1f_{5/2}$	&		&		&		&	-10	&	-47	&	-23	&	10	&	-41	&	-7	&	26	&	-68	&	-21	&	5	&	-40	&	9	\\
$2p_{1/2}$	&	-36	&	-8	&	37	&	-104	&	-34	&	-32	&	-94	&	-36	&	0	&	-84	&	-73	&	-16	&	-92	&	-31	&	24	\\
$2p_{3/2}$	&	-42	&	-6	&	36	&	-108	&	-35	&	-33	&	-99	&	-37	&	-1	&	-88	&	-73	&	-15	&	-97	&	-32	&	24	\\
$1f_{7/2}$	&	90	&	5	&	49	&	14	&	-46	&	-16	&	38	&	-35	&	2	&	53	&	-64	&	-17	&	29	&	-40	&	10	\\
$1d_{3/2}$	&	45	&	9	&	47	&	-48	&	-48	&	-30	&	-28	&	-41	&	-11	&	-16	&	-73	&	-27	&	-34	&	-42	&	6	\\
$2s_{1/2}$	&	-52	&	3	&	40	&	-130	&	-24	&	-43	&	-128	&	-32	&	0	&	-122	&	-73	&	-17	&	-124	&	-25	&	29	\\
$1d_{5/2}$	&	57	&	9	&	46	&	-26	&	-47	&	-26	&	-6	&	-41	&	-8	&	9	&	-70	&	-23	&	-11	&	-40	&	8	\\
$1p_{1/2}$	&	-23	&	11	&	41	&	-97	&	-47	&	-40	&	-79	&	-41	&	-14	&	-70	&	-79	&	-31	&	-82	&	-40	&	7	\\
$1p_{3/2}$	&	-9	&	11	&	43	&	-79	&	-46	&	-35	&	-62	&	-41	&	-12	&	-52	&	-77	&	-29	&	-66	&	-40	&	7	\\
$1s_{1/2}$	&	-120	&	11	&	35	&	-152	&	-40	&	-49	&	-141	&	-39	&	-14	&	-136	&	-84	&	-32	&	-141	&	-37	&	11	\\
\end{tabular}
\end{ruledtabular}
\end{table*}

\begin{table*}[ht!]
\caption{\label{tab2} Numerical results for $\Delta E^T_G$, $\Delta E^G_S$ and $\Delta E^T_F$ (see subsection \ref{test2} for detailed description). The total energy differences, $\Delta E^T_F$ is deduced via Eq.~\eqref{diff} using the data for $\Delta E^S_F$ taken from Ref.~\cite{PhysRevC.63.024312}. All these calculations employ the 2pF function for the charge density. Units are in keV}. 
\begin{ruledtabular}
\begin{tabular}{c|ccc|ccc|ccc|ccc|ccc} 
Orbital	&	$\Delta E^T_G$	&	$\Delta E^G_S$	&	$\Delta E^T_F$	&	$\Delta E^T_G$	&	$\Delta E^G_S$	&	$\Delta E^T_F$	&	$\Delta E^T_G$	&	$\Delta E^G_S$	&	$\Delta E^T_F$	&	$\Delta E^T_G$	&	$\Delta E^G_S$	&	$\Delta E^T_F$	&	$\Delta E^T_G$	&	$\Delta E^G_S$	&	$\Delta E^T_F$	\\
\hline 													
	&	\multicolumn{3}{c|}{$^{16}$O}					&	\multicolumn{3}{c|}{$^{28}$Si}					&	\multicolumn{3}{c|}{$^{32}$S}					&	\multicolumn{3}{c|}{$^{40}$Ca}					&	\multicolumn{3}{c}{$^{48}$Ca}					\\
$2p_{3/2}$	&		&		&		&		&		&		&		&		&		&		&		&		&	22	&	14	&	-128	\\
$1f_{7/2}$	&		&		&		&		&		&		&		&		&		&	87	&	-13	&	-84	&	95	&	-14	&	-77	\\
$1d_{3/2}$	&		&		&		&	34	&	-11	&	225	&	70	&	-16	&	256	&	119	&	-20	&	301	&	126	&	-23	&	305	\\
$2s_{1/2}$	&		&		&		&	-14	&	8	&	188	&	20	&	4	&	218	&	71	&	-3	&	262	&	79	&	-5	&	268	\\
$1d_{5/2}$	&	-82	&	2	&	-254	&	52	&	-16	&	242	&	83	&	-20	&	269	&	125	&	-22	&	309	&	126	&	-23	&	309	\\
$1p_{1/2}$	&	-35	&	-17	&	178	&	81	&	-24	&	432	&	106	&	-25	&	456	&	141	&	-23	&	493	&	141	&	-23	&	493	\\
$1p_{3/2}$	&	-26	&	-21	&	193	&	84	&	-25	&	437	&	107	&	-25	&	460	&	141	&	-24	&	495	&	142	&	-25	&	495	\\
$1s_{1/2}$	&	-5	&	-31	&	395	&	86	&	-23	&	561	&	105	&	-21	&	582	&	135	&	-17	&	616	&	137	&	-18	&	617	\\
\hline		
	&	\multicolumn{3}{c|}{$^{48}$Ni}					&	\multicolumn{3}{c|}{$^{205}$Tl}					&	\multicolumn{3}{c|}{$^{206}$Pb}					&	\multicolumn{3}{c|}{$^{208}$Pb}					&	\multicolumn{3}{c}{$^{209}$Bi}					\\
$1h_{9/2}$	&		&		&		&	278	&	-18	&	136	&	280	&	-18	&	138	&	280	&	-19	&	137	&	281	&	-18	&	139	\\
$3s_{1/2}$	&		&		&		&	228	&	-1	&	366	&	231	&	-1	&	369	&	231	&	-2	&	368	&	232	&	-1	&	370	\\
$2d_{3/2}$	&		&		&		&	247	&	-7	&	382	&	248	&	-6	&	384	&	249	&	-7	&	384	&	250	&	-7	&	385	\\
$1g_{7/2}$	&		&		&		&	289	&	-18	&	427	&	290	&	-18	&	428	&	290	&	-18	&	428	&	291	&	-18	&	429	\\
$2d_{5/2}$	&		&		&		&	244	&	-5	&	395	&	245	&	-5	&	396	&	246	&	-6	&	396	&	247	&	-6	&	397	\\
$1h_{11/2}$	&		&		&		&	271	&	-19	&	393	&	273	&	-19	&	395	&	273	&	-19	&	395	&	275	&	-20	&	396	\\
$1g_{9/2}$	&		&		&		&	283	&	-20	&	530	&	284	&	-19	&	532	&	284	&	-19	&	532	&	286	&	-20	&	533	\\
$1f_{5/2}$	&		&		&		&	294	&	-17	&	572	&	295	&	-16	&	574	&	294	&	-16	&	573	&	296	&	-16	&	575	\\
$2p_{1/2}$	&		&		&		&	263	&	-11	&	528	&	265	&	-10	&	531	&	265	&	-11	&	530	&	267	&	-11	&	532	\\
$2p_{3/2}$	&	74	&	12	&	228	&	261	&	-10	&	537	&	262	&	-9	&	539	&	262	&	-10	&	538	&	264	&	-10	&	540	\\
$1f_{7/2}$	&	148	&	-14	&	309	&	290	&	-18	&	625	&	291	&	-18	&	626	&	291	&	-18	&	626	&	292	&	-18	&	627	\\
$1d_{3/2}$	&	177	&	-20	&	355	&	294	&	-14	&	675	&	296	&	-14	&	677	&	295	&	-14	&	676	&	296	&	-14	&	677	\\
$2s_{1/2}$	&	133	&	-6	&	330	&	272	&	-11	&	643	&	273	&	-11	&	644	&	272	&	-11	&	643	&	274	&	-11	&	645	\\
$1d_{5/2}$	&	176	&	-21	&	479	&	293	&	-15	&	699	&	295	&	-16	&	700	&	294	&	-15	&	700	&	296	&	-16	&	701	\\
$1p_{1/2}$	&	189	&	-20	&	554	&	291	&	-10	&	750	&	293	&	-10	&	752	&	292	&	-10	&	751	&	294	&	-10	&	753	\\
$1p_{3/2}$	&	188	&	-21	&	594	&	292	&	-12	&	757	&	294	&	-12	&	759	&	292	&	-11	&	758	&	293	&	-11	&	759	\\
$1s_{1/2}$	&	183	&	-14	&	676	&	285	&	-8	&	797	&	287	&	-8	&	799	&	286	&	-7	&	799	&	287	&	-7	&	800	\\
\end{tabular}
\end{ruledtabular}
\end{table*}

\end{document}